\documentclass[10pt, conference, compsocconf]{IEEEtran}

\input{./fig_margin_patch_code.tex}

\ifCLASSOPTIONcompsoc
    \usepackage[caption=false, font=normalsize, labelfont=sf, textfont=sf]{subfig}
\else
\usepackage[caption=false, font=footnotesize]{subfig}
\fi

\usepackage{graphicx}

\ifCLASSINFOpdf
\else
\fi
\usepackage{balance}

\IEEEoverridecommandlockouts 
\usepackage{booktabs}
\usepackage{xcolor}
\usepackage{tikz,pgf,pgfplots}
\usepgfplotslibrary{statistics}
\pgfplotsset{compat=1.14}
\usepackage{graphicx}
\usepackage[left=0.76in,right=0.85in,top=1.02in,bottom=0.98in]{geometry}
\usepackage{listings}
\usepackage[OT4,T1]{fontenc}
\usepackage{caption}
\usepackage[cmex10]{amsmath}
\usepackage{amsthm,amssymb}

\usepackage{url}
\usepackage{multirow}
\usepackage{url}
\usepackage{datapie}
\usepackage{subfig}
\pgfcreateplotcyclelist{\mylist}{
	{black,mark=none,thick}, 
	{black,mark=none,thick,densely dashed},
	{black,mark=none,thick,densely dotted},
	{red,mark=none,thick}, 
	{red,mark=none,thick,densely dashed},
	{red,mark=none,thick,densely dotted},
	{dash pattern=on 8pt off 2 pt on 2pt off 2pt,thick,mark=none}, 
}
\pgfcreateplotcyclelist{\mylistmarks}{
	{black,mark=*}, 
	{black,mark=square,densely dashed,mark options=solid},
	{black,mark=triangle,densely dotted,mark options=solid},
	{mark=o,dash pattern=on 8pt off 2 pt on 2pt off 2pt,mark options=solid}, 
}

\usepackage[left=0.76in,right=0.85in,top=1.02in,bottom=0.98in]{geometry}

\begin{document}
%
\title{\Large \textbf{A Conceptual Framework for Assessing Anonymization-Utility Trade-Offs Based on Principal Component Analysis}}
\author{

\author{
    \IEEEauthorblockN{Giuseppe D'Acquisto\IEEEauthorrefmark{1}, Maurizio Naldi\IEEEauthorrefmark{1}\IEEEauthorrefmark{2}}
    \IEEEauthorblockA{\IEEEauthorrefmark{1}Dept. of Civil Engineering and Computer Science\\University of Rome Tor Vergata\\
Via del Politecnico 1, 00133 Rome, Italy\\dacquisto@ing.uniroma2.it, maurizio.naldi@uniroma2.it}
    \IEEEauthorblockA{\IEEEauthorrefmark{2}Dept. of  Law, Economics, Politics and Modern languages\\LUMSA University\\Via Marcantonio Colonna 19, 00192 Rome, Italy\\m.naldi@lumsa.it}
}

}

%


\maketitle

\begin{abstract}
An anonymization technique for databases is proposed that employs Principal Component Analysis. The technique aims at releasing the least possible amount of information, while preserving the utility of the data released in response to queries. The general scheme is described, and alternative metrics are proposed to assess utility, based respectively on matrix norms; correlation coefficients; divergence measures, and quality indices of database images. This approach allows to properly measure  the utility of output data and incorporate that measure in the anonymization method. 
\end{abstract}


%
\IEEEpeerreviewmaketitle

\section{Introduction}

\IEEEoverridecommandlockouts\IEEEPARstart{A}{nonymization} may be employed in databases to achieve privacy protection. Through anonymization, personally identifiable information is removed, or obfuscated so that the people whom the data describe remain anonymous. Several techniques have been proposed in the literature to achieve anonymization (see, e.g., \cite{domingo2016database} for a survey).

While those techniques may exhibit several degrees of robustness against re-identification attacks (see, e.g., Bayesian attacks for differential privacy schemes  \cite{naldi2014differential,zhangstatistical}, or attacks based on the exploitation of externalities \cite{naldi2017mr}, or the use of recolouring techniques borrowed from signal processing \cite{naldi2018hiding}), they often fail to take into account the resulting utility of released data. For example, in randomization-based techniques,  the amount of added noise needed to obscure the true data may be so large as to make the query output useless. In differential privacy, this may require a careful choice of the mechanism parameters \cite{naldi2015differential,kohli2018epsilon}. The contrasting wishes to minimize the risk of re-identification while providing useful data to queriers gives rise to an anonymization-utility trade-off, which triggers the need for proper utility measures and utility-aware anonymization techniques \cite{domingo2017methodology}. 

In this paper, we introduce a technique that would provide a utility-aware anonymization for databases. Our technique is based on a novel use of Principal Component Analysis (PCA), a statistical technique to achieve dimensionality reduction that has already been employed to protect privacy \cite{dwork2014analyze}. The  technique is typically employed in signal processing applications to reduce the information as little as possible, while maintaining the quality of the output data. Instead, here we strive to achieve the maximum possible loss of information (and then potential individuals' identification), while preserving the utility of the output data. For that purpose, we introduce metrics of utility and incorporate them in the anonymization technique. The new approach we propose can therefore boast a built-in utility-aware mechanism, in contrast to existing anonymization technique, where data utility is not explicitly considered or is not properly measured.

Here we do not fully explore the technique, but rather set up a conceptual framework and propose several alternatives for its implementation. Our contributions are:
\begin{itemize}
    \item incorporating utility measurement in the proposed PCA-based anonymization scheme (Section \ref{sec:anon}).
    \item proposing four categories of metrics to assess the residual utility of released data (Section \ref{sec:metrics});
\end{itemize}

\section{Principal Component Analysis}
\label{sec:pca}
Principal Component Analysis (PCA) is a statistical technique that applies linear transformations to a set of variables so as to arrive at a new set of variables that are statistically uncorrelated. It is typically employed to reduce the number of variables (the principal components, i.e. a subset of the new variables mentioned above) that retain most of the information contained in the original set. An extensive treatment of PCA is contained, e.g., in \cite{jolliffe2011principal}

The PCA technique considers a set of $n$ observations of $p$ variables. Each observation represents a point in $\mathbb{R}^p$. Actually, we are dealing with databases, where data concerning subjects may be seen as matrices where each row is a record representing the data concerning a specific individual, and the columns are the fields representing the attributes of the data subject. For the time being, we consider just databases made of numerical entries: we defer the case of categorical attributes to a later implementation of our method.

As recalled, each principal component is actually a linear transformation of the original variables. We can have at most the same number of principal components as the original variables, i.e. $p$, but the main aim of PCA is to reduce the dimensionality of the description: we wish to retain $m$ principal components, with $m\le p$. 

\begin{figure}
    \centering
    \includegraphics[width=0.8\columnwidth]{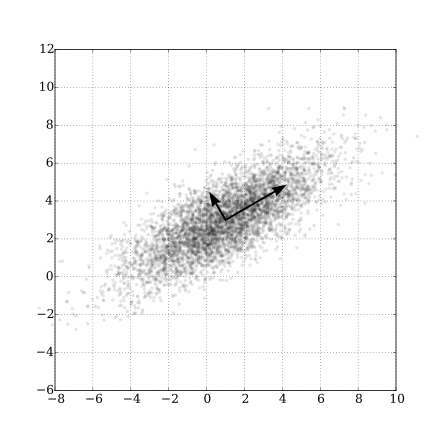}
    \caption{Example of PCA for a multivariate Gaussian dataset}
    \label{fig:pca}
\end{figure}

The principal components are built starting from the most important one, i.e., that retaining the major portion of the information. In mathematical terms, that means building a linear combination of the $p$ variables (i.e., setting its weights) so that it exhibits the maximum variance (under the constraint that the weight vector has unitary length). A graphical representation of PCA in $\mathbb{R}^{2}$ is shown in \figurename~\ref{fig:pca}, where the two axes representing the two principal components are shown for data following a multivariate Gaussian distribution.

The PCA technique has already been employed as an exploratory tool to describe large databases \cite{meglen1992examining,emberly2004flexibility}. Though  the traditional way to employ PCA is to achieve dimensionality reduction by sacrificing as little information as possible (i.e., removing principal components starting from the least ones), for our purposes we wish instead to apply the PCA technique in a different way, to achieve as much information reduction as possible.  

\section{Anonymization by PCA}
\label{sec:anon}
After describing how PCA works, we now turn to its use to anonymize a database. In this section, we describe the overall method, starting with the rationale and then providing the step-by-step procedure.  

PCA defines a new set of axes that represent linear combinations of the original variables. The typical application of PCA includes removing some principal components, by retaining the largest ones. Such a procedure has two limitations for our purpose:
\begin{itemize}
    \item removing the least principal components achieves a reduction of information as small as possible;
    \item the principal components do not represent physically meaningful variables (e.g. a linear combination of height and income).
\end{itemize}   
We overcome those limitations through the following modification of the typical application:
\begin{itemize}
    \item we start removing principal components starting from the largest one;
    \item after removing the largest principal components, we project the resulting data onto the original axes, so that we revert to the original attributes of the subjects (e.g., height and income separately). 
\end{itemize}

The resulting procedure goes through the following steps:
\begin{enumerate}
    \item Form an array corresponding to the full database, represented as a collection of vectors in $\mathbb{R}^p$;
    \item Apply a PCA transformation to the array;
    \item Remove the principal component corresponding to the largest eigenvalue;
    \item Project the resulting collection of vectors on the original set of axes in $\mathbb{R}^p$;
    \item Apply a utility metric;
    \item If the utility is still large enough then go back to step 2;
    \item Transform the collection of vectors back into a database.
\end{enumerate}

Steps 1-4 and 7 are quite straightforward and anyway well described in several textbooks. Instead, the choice of utility metrics and its employment are dealt with in the next section. The check embodied by Step 6 allows us to see if further information can be removed: if the utility is still large enough, that means that we can further reduce it and achieve a higher degree of anonymization.  

\section{Utility metrics}
\label{sec:metrics}
In the procedure we have sketched in Section \ref{sec:anon}, we have mentioned the use of utility metrics. Those metrics allow us to assess how far the anonymized version of the database is from the true database or, in other terms, if the anonymized version can still be useful for queriers. In Section \ref{sec:anon} we have not provided further details about such metrics. We fill the gap in this section, by providing four different proposals, i.e., four metrics that can serve that purpose. Those metrics are not intended to be alternative, since they could be employed in conjunction.

The four metrics (or classes of metrics, since we may have some alternative possibilities for each of them) we propose are:
\begin{enumerate}
    \item Matrix norms;
    \item Correlation;
    \item Divergence measure;
    \item Database image quality.
\end{enumerate}

\textbf{Matrix norms.} The first metric is based on the matrix representation of the database. Our database of $n$ records and $p$ fields is represented as a matrix with $n$ rows and $p$ columns. In the following, we indicate the number of columns by $m$, since we assume to act after PC removal. We can first apply a standardisation procedure, so that the results are not influenced by the different ranges of the variables: for example, if the two fields were \textit{height} and \textit{income}, measured respectively in meters and euros, the differences in the income field would dominate. If we call $A$ the matrix representing the true database and $B$ the matrix representing the anonymised database (with elements $a_{ij}$ and $b_{ij}$ respectively), the standardisation field by field would require the computation of the mean and the standard deviation for each field
\begin{equation}
\begin{split}
    \mu_{j} &= \frac{1}{n}\sum_{i=1}^{n}a_{ij}\\
    \sigma_{j}&= \sqrt{\frac{\sum_{i=1}^{n}( a_{ij}-\mu_{j})^{2}}{n}}.
\end{split}
\end{equation}

By acting on the standardized values $\hat{a}_{ij}=\frac{a_{ij}-\mu_{j}}{\sigma_{j}}$ and $\hat{b}_{ij}=\frac{b_{ij}-\mu_{j}}{\sigma_{j}}$, we form the difference matrix $D$, whose generic element is $d_{ij} = \vert \hat{a}_{ij}- \hat{b}_{ij} \vert = \frac{\vert a_{ij}-b_{j}\vert}{\sigma_{j}}$, $1\le i \le n$ and $1 \le j \le m$ (we are not interested in the sign of the distance between the true database and the anonymized one).

Having defined the difference matrix, we can now adopt a metric $M$ for the size of its contents. Among the most widespread metrics for this purpose, we can consider for example the sum of all its elements:
\begin{equation}
    M = \sum_{i=1}^{n}\sum_{j=1}^{m} d_{ij}.
\end{equation}
Alternatively, we can consider one of the norms of $D$ (see Chapter 2.6.5 of \cite{zwillinger2002crc}), e.g., the $L_1$ norm
\begin{equation}
    M = \underset{1\le i \le n}{\operatorname{max}} \sum_{j=1}^{m} d_{ij},
\end{equation}
which we have adapted to our case, since we are more interested in picking the record showing the largest distance from its true value (hence summing along columns rather than along rows as in the textbook formulation of that norm).
Another norm of possible use is the Frobenius norm
\begin{equation}
    M = \sqrt{\sum_{i=1}^{n}\sum_{j=1}^{m}d_{ij}^{2}}
\end{equation}

Whatever the norm we employ, the larger it is, the less useful the anonymized database is.

\textbf{Correlation.} The second metric we consider is correlation. Though a correlation measure is not defined for matrices, we can convert the matrices of interest into vectors, by reading matrices row-by-row, and then apply the usual correlation definition. In our case, since we have two $n \times p$ matrices $A$ and $B$, representing respectively the true database and the PCA-treated one, we can convert them into two vectors $v$ and $w$ of length $nm$, by defining a row index $r$ and a column index $c$ and relating them to the vector index $k$ as follows
\begin{equation}
\begin{split}
    r &= \left\lceil \frac{k-1}{m} \right\rceil +1\\
    c &= k-(r-1)m.
\end{split}
\end{equation}
We can then apply the classical definition, so that the correlation is
\begin{equation}
    \rho= \frac{\sum_{i=1}^{nm}(v_{i}-\overline{v})(w_{i}-\overline{w})}{\sqrt{\sum_{i=1}^{nm}(v_{i}-\overline{v})^{2}\sum_{i=1}^{nm}(w_{i}-\overline{w})^{2}}}
\end{equation}

Contrary to the use of the norm metric, now the larger this correlation coefficient, the more useful the anonymized database is.

\textbf{Divergence measure.} The values observed for the $m$ fields of the $n$ records can be considered as a realization of a multivariate random variable. When comparing the two databases (the true one and the anonymized one), we can then compare the two associated multivariate random variables. 

We can then adopt a measure of distance between the two distributions as a utility metric for the anonymization process. A possible metric is Kullback-Leibler (KL) divergence \cite{kullback1951information}. For our case, we have the two multivariate random variables $a$ and $b$, pertaining respectively to the true database and the anonymized one, with probability density function $\phi(a)$ and $\psi(b)$ respectively, so that the KL divergence is
\begin{equation}
D_{KL} = \int_{a}\phi(x)\log \frac{\phi(x)}{\psi(x)}dx.
\end{equation}
Though this is not properly a metric (it is not symmetric), we are going to employ it by keeping the true database as a reference, i.e., we always apply it in the same direction, so that the lack of symmetry is not relevant. 

If we assume that the two random variables follow a multivariate normal distribution, the KL divergence takes the form \cite{kakizawa1998discrimination,runnalls2007kullback}:
\begin{equation}
\label{eq:kl}
\begin{split}
    D_{KL} = &\frac{1}{2}\left[ \mathrm{tr}(R_{a}R_{b}^{-1}) - \log \frac{\vert R_{a} \vert}{\vert R_{b} \vert} -m\right]\\& + \frac{1}{2}\left[(\Bar{a}-\Bar{b})^{T}R_{b}^{-1}(\Bar{a}-\Bar{b})\right],
\end{split}
\end{equation}
where $R_{a}$ and $R_{b}$ are the autocovariance matrices of the two variates $a$ and $b$. 

Under this assumption, the KL divergence can be computed by estimating the autocovariance matrices and plugging them into Equation (\ref{eq:kl}). The larger the divergence is, the less useful is the anonymized database.

\textbf{Database image quality.} In order to assess how much utility is retained after principal components are removed, we can resort to the representation of a database as an image. In fact, our database can be viewed as a color-scale image, where the level of color of the pixel of indices $(x,y)$ is the value of the attribute $y$ of the record $x$ (the same procedure applies to a greyscale image). 

We can liken anonymization to image compression. Image compression aims at reducing the bit load associated to the image as much as possible while preserving the quality of the image, Reducing the bit load amounts to reducing the information content of the image, which is what we wish to achieve through anonymization. Since PCA has been applied to image compression as well \cite{du2007hyperspectral,lim2001principal,wang2006independent}, we can exploit the representation of databases as images to translate the results of image compression through PCA into database transformation through PCA as well.

Utility reduction can therefore be assessed through the tools that have been devised to assess the quality of an image after compression. Assessing the quality of an image is intrinsically difficult \cite{wang2002image}, and several approaches have been proposed, which may be classified into the two classes named subjective and objective approaches. 

A major index employed under the subjective approach is the Mean Opinion Score (MOS). The MOS has first been conceived for audio signals \cite{itu1999subjective}, and then extended to images and videos \cite{series2012methodology}. Several possibilities exist for the testing methods (see, e.g., \cite{pinson2003comparing} for a comparison). The differences are mainly linked to the number of images that are shown to the human observer and the grading scale. As to to the first issue, the observer may be presented either just with the image whose quality is to be assessed (single stimulus approach) or with two images, i.e., the image under test and its reference (free of impairments). In the latter case we have the so-called double stimulus approach. As to grading, each observer can assign a mark to the image under test, by employing either a discrete scale (e.g., a five points scale) or a continuous scale (which means indicating a point on a line, where notches are shown corresponding to marks from 0 to 100). The overall procedure therefore goes through the following steps:
\begin{enumerate}
    \item A set of human observers is selected, following the principles of experiment design;
    \item each observer is presented with the image(s) to be assessed;
    \item each observer assigns a grade on the scale indicated;
    \item the arithmetic average of the grades assigned by the observers is computed to provide the Mean Opinion Score. 
\end{enumerate}
Despite its limitations \cite{streijl2016mean}, MOS remains a well established tool to assess the quality of images, and is considered as the ground truth assessment in many contexts, e.g. when assessing retargeting tasks \cite{ma2012image}. In our case, we envisage opting for a double stimulus approach, with a continuous scale to avoid the quantization errors typically associated to the discrete scale. 

Under the objective approach, two major indices are instead the Peak Signal-to-Noise Ratio (PSNR) and the Structural Similarity Index (SSIM) \cite{hore2010image}. For an image of size $n\times m$ pixels, and a resolution of 8 bit/pixel, the PSNR for the reference image $f$ and the (distorted) image $g$ under test is
\begin{equation}
    \mathrm{PSNR} = 10\log_{10} \frac{255^2}{\mathrm{MSE}(f,g)},
\end{equation}
where the Mean Squared Error MSE is defined as
\begin{equation}
    \mathrm{MSE}(f,g) = \frac{1}{nm}\sum_{i=1}^{n}\sum_{j=1}^{m} (f_{ij}-g_{ij})^{2}.
\end{equation}
The PSNR is always non negative but is not bounded; a high PSNR means that the database with image $g$ is close to the database with image $f$ (since high PSNR implies a low MSE).
Instead, SSIM is defined as the product of three quality reduction factors, namely loss of correlation $s$, luminance distortion $l$, and contrast distortion $c$:
\begin{equation}
\begin{split}
    \mathrm{SSIM} &= s(f,g)l(f,g)c(f,g)\\
    &=\frac{\sigma_{fg}+C_{3}}{\sigma_{f}\sigma_{g}+C_{3}}\times\frac{2\mu_{f}\mu_{g}+C_{1}}{\mu_{f}^{2}+\mu_{g}^{2}+C_{1}}\times\frac{2\sigma_{f}\sigma_{g}+C_{2}}{\sigma_{f}^{2}+\sigma_{g}^{2}+C_{2}},
\end{split}
\end{equation}
where $\mu_{f}$ and $\mu_{g}$ are the average luminances of $f$ and $g$, $\sigma_{f}$ and $\sigma_{g}$ the respective standard deviations, and $C_{1}$, $C_{2}$ and $C_{3}$ are three positive constants to avoid having a null denominator. In contrast to the PSNR, SSIM is bounded and takes value in the [0,1] range, with higher values representing images $g$ closer to the reference image $f$.

Though adopting the image paradigm, we must however stop short of applying a full parallelism when employing PCA. In fact, images are typically correlated both horizontally and vertically. While we may assume that a horizontal correlation (among the attributes of a single record) exists in databases, it would be hazardous to assume that a vertical correlation (between different records) exists as well. 

A possible approach would therefore consist in applying image compression through PCA to a set of well-known references pictures, e.g., Lena \cite{munson1996note}, by first removing the vertical correlation (e.g., by randomly shuffling the image rows) and then progressively removing the largest principal components, assessing the resulting image quality through MOS, and proceeding till the image is still recognizable. We would then consider the number of removed principal components as the limit number of principal components that we can safely remove when applying PCA to a database.

We expect the results to depend on the particular image we are considering, through the structure of eigenvalues. In order to reduce that bias, we envisage building sets of image libraries that exhibit a similar eigenvalues structure, e.g. by looking at the dominant eigenvalue or a parameter of the eigenvalue curve. Some sets would allow for just a limited number of components to be removed before the image is not recognizable, while other sets would allow for a more extensive removal before the degradation is excessive. This could be related to the identifiability of the processed image (and, in parallel, of the database to be anonymized), and be considered for properly training the MOS-based anonymizer.

\begin{figure}
    \centering
    \includegraphics[width=0.8\linewidth]{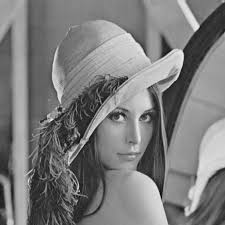}
    \caption{Lena original image}
    \label{fig:lena-original}
\end{figure}

We can report now an early example of the effect of principal component removal. In \figurename~\ref{fig:lena-original}, we show the original picture of Lena in greyscale.

\begin{figure} 
    \centering
  \subfloat[Removal of largest principal component\label{1a}]{%
       \includegraphics[width=0.8\linewidth]{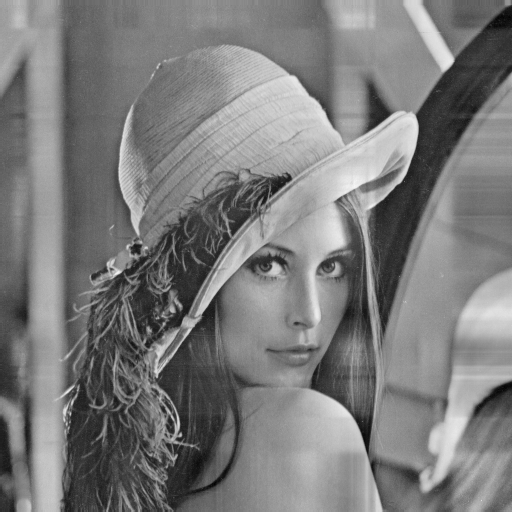}}
    \hfill
  \subfloat[Removal of the two largest principal components\label{1b}]{%
        \includegraphics[width=0.8\linewidth]{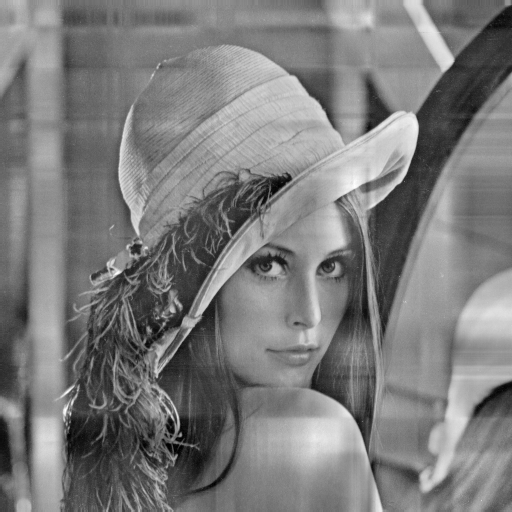}}
    \\
  \subfloat[Removal of the five largest principal components\label{1c}]{%
        \includegraphics[width=0.8\linewidth]{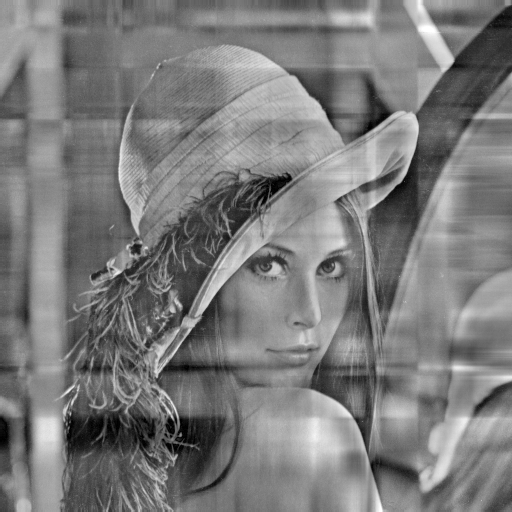}}
  \caption{Impact of principal component removal.}
  \label{fig:lenapca} 
\end{figure}

In \figurename~\ref{fig:lenapca}, we report the image as it appears after removing respectively just the largest, the two largest, or the five largest principal components. The removal of the largest PC does not impact the quality, while we see some artifact when we remove two PCs, and the artifacts get more significant when we remove 5 PCs, though the image is still well recognizable.

The resulting objective quality indices for those cases are shown in Table \ref{tab:removal}: similarly to what a quick visual analysis of the picture would tell us, those indices fall sharply as more PCs are removed.

\begin{table}[]
    \centering
    \begin{tabular}{ccc}
    \toprule
    No. of component removed     & SSIM & PSNR  \\
    \midrule
    1   & 0.9335 & 20.0081\\ 
    2 & 0.9036 & 17.9902\\
    5 & 0.8614 &16.1854\\
    \bottomrule
    \end{tabular}
    \caption{Quality indices after principal components removal}
    \label{tab:removal}
\end{table}

\begin{figure}
    \centering
\begin{tikzpicture}
	\begin{axis}[
		xlabel={Eigenvalue order},
		ylabel={Eigenvalue},
		xmin=1,
		xmax=5,
	]
	\addplot[mark=none] {29624 + 389905/(1+(x/1.7664)^(2.2111))};
	\addplot[mark=*,color=blue,only marks] coordinates{(1,333320) (2,197204) (3,124780) (4,80285) (5,67232)};
	\end{axis}
\end{tikzpicture}    
\caption{Decay of dominant eigenvalues for Lena}
    \label{fig:eigenstructure}
\end{figure}
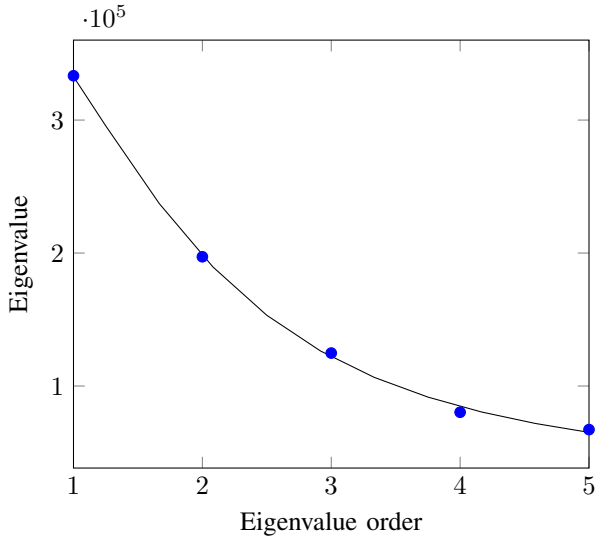

As to the identifiability properties of the image as revealed by the eigenvalue structure, we show in \figurename~\ref{fig:eigenstructure} how the eigenvalues decay. The superimposed curve is the best fit ($R^{2}=0.9993$) obtained with a symmetrical sigmoidal curve
\begin{equation}
    y = d + \frac{a-d}{1+\left( \frac{x}{c}\right)^b},
\end{equation}
where $c$ and $b$ respectively act as a scale factor and a shape factor, with the latter suitable to represent the speed at which eigenvalues decay. In our case, we have $b=2.21$.

\section{Conclusions}

Previous proposals to anonymize data do not explicitly take into account the residual utility of data after the database has been processed to achieve privacy. We propose enhancing an approach based on Principal Component Analysis by incorporating a measure of utility and proposing four classes of metrics that can be used for that purpose. This approach represents a step forward in the direction of devising reliable privacy protection mechanisms that actually provide usable data. 

\section*{Acknowledgments}
The authors wish to thank Dr. Fabio Ricciato of Eurostat for the fruitful discussions on the subject, and Mr. Luca Della Gatta for contributing \figurename~\ref{fig:lenapca} and Table \ref{tab:removal}.

%
%






%


\newpage

\end{document}